# Mathematical modeling of the treatment response of resection plus combined chemotherapy and different types of radiation therapy in a glioblastoma patient

**Farshad Samadifam[a]. Ehsan Ghafourian[b]**
a. Department of Biomedical Engineering, Medical College of Wisconsin, Milwaukee, Wisconsin, USA
b. Department of Computer Science, Iowa State University, Ames, IA, USA

## Abstract

**Background** Gliomas are the most common brain tumors, which vary from low to high-grade. Despite the improvements in imaging techniques in the last decade, all parts of the tumor are not visible in these images, due to having a threshold of detection often lower than most of the tumor.
**Method** We simulated a brain domain to model heterogeneous growth plus gross total resection following concurrent chemo- and radiation therapy and adjuvant chemotherapy using temozolomide for a patient, diagnosed with glioblastoma, for different radiobiological alpha parameter as high, medium, and low sensitivity. The extent of difference in the result of treatment, area of the tumor, and survival time relating to those sensitivity parameters were shown. Moreover, based on the threshold of detection of MRI (T1Gd), the area of the tumor visible to this imaging technique was simulated. Furthermore, the extent of difference in the result of treatment for both conventional and hyperfractionated radiotherapy was investigated.
**Conclusion** The survival time of patient was increased by 2 months and more than 3 months at the lowest and highest sensitivity of radiation therapy, respectively.The result showed hyperfractionated radiotherapy was more effective than conventional radiotherapy in reducing the concentration of cancerous cells,as hyperfractionated method allows administration of higher total dose.

**Keywords:** Chemotherapy; Glioblastoma; Magnetic Resonance Imaging; Mathematical modeling; Radiotherapy; Resection

## 1. Introduction

Glioblastomas are among the most aggressive brain tumors in which invasiveness and high diffusivity are two characteristics of these tumors. Glioblastoma multiforme (GBM) is one of these tumors with survival time of at most 15 months [1]. The most common therapies for these tumors are surgery, chemotherapy,and radiotherapy.

Despite the improvement in the field of tumor imaging, all parts of the tumor are not visible in these images, due to having a specific threshold of detection which doesn't cover some parts of the tumor that are below that threshold. This problem clearly highlights the need for modeling the treatment of these tumors and their growth more accurately and with greater precision. Mathematical modeling is a great way of dealing with the biological problems that allow us to have insight into solutions for these problems. Mathematical modeling of glioma growth was first modeled by Tracqui et al. based on proliferation and diffusion rate [2]. Similarly, Woodward et al. studied diffusion and growth rate in gliomas with resection as the only treatment [3]. Swanson et al. extended a mathematical model, known as reaction-diffusion, based on proliferation and diffusion rate to study their effects in white and grey matter. They presumed that the diffusion rate in white matter is five times greater than the grey matter. The radius of the tumor at detection and death considered to be 1.5 and 3 cm respectively [4]. Other similar models in this field were reviewed by Swanson et al[5].Regarding chemotherapy drugs, Parney et al. investigated several chemotherapeutic agents and believed that overall survival for glioblastomas' patients is higher when temozolomide was applied rather than other agents such as procarbazine. Moreover, temozolomide proved to have less toxicity than other drugs [6]. Stupp et al. investigated radiotherapy plus temozolomide for 573 glioblastoma patients from 85 centers and concluded that temozolomide increased the survival time of patients with minimal toxicity [7]. Radiation therapy was modeled based on the reaction-diffusion equation of Swanson by Rockne et al. The radiation effect was added to the previous reaction-diffusion equation in order to account for the loss of cancerous cells due to radiotherapy [8]. Hathout et al. applied the same diffusion-reaction equation and investigated the extent of surgical resection for glioblastoma [9]. Moshtaghi-Kashanian et al simulated glioblastoma growth followed by 100%, 105%,and 125% resection and radiotherapy for different locations [10]. Ghafourian et al investigated an ensemble model for the diagnosis of brain tumors through MRIs [11].

In this study a spatio-temporal growth of inner tumor up to the resection time was modeled, then concurrent heterogeneous chemotherapy using temozolomide and radiotherapy plus adjuvant chemotherapy was simulated for different alpha parameters as high, medium, and low radiosensitivity. The diffusion and chemotherapy killing rate assumed to be different in grey and white matter. Moreover, the area of the tumor visible to MRI(T1Gd), and survival time has been simulated.

Furthermore, the effect of conventional and hyperfractionated radiotherapy on the concentration of cancerous cells and the area of the tumor were shown.

## 2. Material and Methods

We have selected one patient data among 32 patients whose clinical data were reported by Wang [12]. This patient had undergone gross total resection (GTR), chemotherapy using temozolomide (TMZ), and radiotherapy. All data are outlined in Table 1.

**Table 1** Patient data

| Age | Sex | EOR | RPA class | T1Gd radius | $D^1$ | $\rho^2$ | $K^3$ |
|---|---|---|---|---|---|---|---|
| 53 | M | GTR | IV | 9.7 mm | $28.6 \frac{mm^2}{year}$ | $8.2 \frac{1}{year}$ | $2.55 \frac{cell}{m^2}$ |

We used image-processing techniques to model brain geometry in COMSOL Multiphysics 5.4 , to accurately distinguish between the grey and white matter. The geometry of the brain and tumor is depicted in Fig 1. The grey matter, white matter, inner tumor, and resected area in Figure 1 are shown in grey, white, orange, and bright orange respectively. For the sake of simplicity, it was assumed that the geometry of the tumor is spherical with a T1Gd radius size reported from Wang data [12]. As this patient had undergone gross total resection, we have modeled tumor growth from its 60% radius in order to have a heterogeneous distribution of cancerous cells in geometry before modeling surgery.

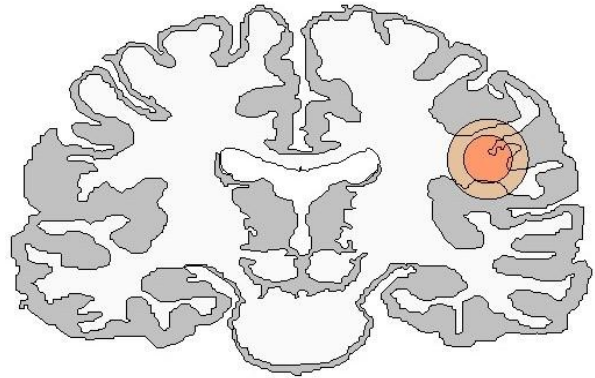

Figure 1. Brain geometry including white and grey matter, initial tumor, and resected area

J.D.Murray in the early 1990s formulated growth of an infiltrating glioma as a conservation equation as follows [13]:

Rate of change of tumor cell population = diffusion(motility) of tumor cells + net proliferation of tumor cells (1)

They described the equation (1) mathematically as follows in equation (2):
$$\frac{\partial c}{\partial t} = \nabla . J + \rho c \qquad (2)$$
The above equation under the assumption of classical gradient-driven Fickian diffusion turns into equation (3):
$$\frac{\partial c}{\partial t} = \nabla . (D \nabla c) + \rho c \qquad (3)$$
c(x,t) represent the concentration of tumor cells at location x and time t, ρ is the net proliferation rate, and D(x) is the diffusion coefficient at location x representing active motility of cells which is different in white and grey matter. The migration rate is generally faster in the white matter than in the grey matter, so a five-fold difference was used: $D_w = 5\, D_g$ [14]. Moreover, equation (3) mathematically indicates the growth of the tumor as a traveling wave, with a velocity of v, expanding radially in accordance with Fisher's approximation $v = \sqrt{2D\rho}$ [5]. Swanson reformulated equation (3) in order to put a limitation in the proliferation of the tumor by introducing the carrying capacity of tissue (K), Which is the maximum capacity of tumor cells in the tissue. The aforementioned equation (4) is as follows:

$$\frac{\partial c}{\partial t} = \nabla.(D(x)\nabla c) + \rho c \left(1 - \frac{c}{k}\right) \quad (4)$$

---

[1] Diffusion rate

[2] Proliferation rate

[3] Carrying capacity

Swanson et al investigated the effect of chemotherapy on gliomas by adding a loss term (G) to the previous proliferation-invasion model. They wrote the new formula as [14]:

$$\frac{\partial c}{\partial t} = \nabla.(D(x)\nabla c) + \rho c \left(1 - \frac{c}{k}\right) - G(x,t)c \quad (5)$$

$$G(x,t) = \begin{cases} 0 & \text{for } t \notin \text{therapy} \\ \begin{cases} K_w & \text{for white matter} \\ K_g & \text{for grey mater} \end{cases} & \text{for } t \in \text{therapy} \end{cases} \quad (6)$$

They expected that the effect of chemotherapy drug to be much less than that of grey matter, as tumor cells diffusion in grey matter is slower than white matter, so they spend more time in grey matter and the likelihood of being exposed to the drug is higher [16]. We have used temozolomide as the chemotherapy drug as it has been believed to be effective in treating recurrent gliomas. Moreover, Temozolomide is more tolerable for elderly patients and its toxicity is relatively low compared to other chemotherapy drugs [17]. Powathil et al. calculated the average killing rate of temozolomide to be $0.0196 \frac{1}{day}$ [18].

Rockne et al. [18] described the effect of radiotherapy by adding a loss term (R), due to the radiotherapy effect, to equation (5). Powathil et al. utilized the widely common linear-quadratic radiobiological model that considers the effects of space,time,and different fractionations per day . The extension to the previous formula is as follows[18]:

$$\frac{\partial c}{\partial t} = \nabla.(D(x)\nabla c) + \rho c \left(1 - \frac{c}{k}\right) - R(x,t,Dose)c \quad (7)$$

$$R = R_{eff} K_R(t) \quad (8)$$

$$R_{eff} = \alpha\, n\, Dose + \beta\, (n\, Dose)^2 \left[g(\mu\tau) + 2\left(\frac{\cosh(\mu\tau) - 1}{(\mu\tau)^2}\right) h_n(\varphi)\right] \quad (9)$$

$$g(\mu\tau) = 2\left(\frac{\mu\tau - 1 + exp\,(-\mu\tau)}{(\mu\tau)^2}\right) \quad (10)$$

$$h_n(\varphi) = 2\left(\frac{n\varphi - n\varphi^2 - \varphi + \varphi^{n+1}}{n(1 - \varphi)^2}\right) \quad (11)$$

$$\varphi = \exp(-\mu(\tau + \Delta\tau)) \quad (12)$$

$$\mu = \frac{Ln2}{T_h} \quad (13)$$

$$K_R(t) = \begin{cases} 1 & t \in Therapy \\ 0 & t \notin Therapy \end{cases} \quad (14)$$

Equation (9) shows the relation of effective dose ($R_{eff}$) to radiation dose in units of Gy (Gy=1J/Kg), and radiobiological parameters $\alpha\left(\frac{1}{Gy}\right)$ and $\beta\left(\frac{1}{Gy^2}\right)$.Rockne et al. held the ratio of $\frac{\alpha}{\beta} = 10$. $\alpha$ known as radiation sensitivity parameter of radiotherapy is different depending on the patient. Large $\alpha$ shows high sensitivity to radiation therapy, while small $\alpha$ shows low sensitivity [20].The duration of irradiation for both conventional and hyperfractionated radiotherapy was considered to be 12 minutes.Time interval of 1 day and 6 hours was used for conventional and hyperfractionated radiotherapy respectively.

The extension of all aforementioned formulas, in order to describe the effects of radiotherapy and chemotherapy, is as follows:

$$\frac{\partial c}{\partial t} = \nabla \cdot (D(x)\nabla c) + \rho c \left(1 - \frac{c}{k}\right) - R(x,t,Dose)c\left(1 - \frac{c}{k}\right) - G(x,t)c\left(1 - \frac{c}{k}\right) \quad (15)$$

The initial and boundary condition for equation (15) is as follows:

$$initial\ condition\ c(x,0) = c_0(x)$$

$$boundary\ condition\quad n.\nabla c = 0\ on\ \partial B\ (the\ brain\ boundary)$$

The above boundary condition makes cancerous cells not leave the brain.

## 3. Results and Discussion

### 3.1 Investigating the treatment for low, medium, and high radiosensitivity parameter

Firstly, the growth of the tumor with 60% of the resected radius was simulated until reaching the resected area. The time, taken to grow from 60% of its resected radius to the resected radius, was calculated using fishers' approximation:

$$t = \frac{r_{resection} - r_{60\%}}{v} \quad (16)$$
$$v = 2\sqrt{\rho D} \quad (17)$$

Reaching the resected size, the tumor cell density inside the resected area was set to zero to simulate gross total resection and heterogenous distribution of cancerous cells around resected area was used as an initial condition for the growth of tumor after surgery. Distribution of cancerous cells after surgery is depicted in Figure 2.

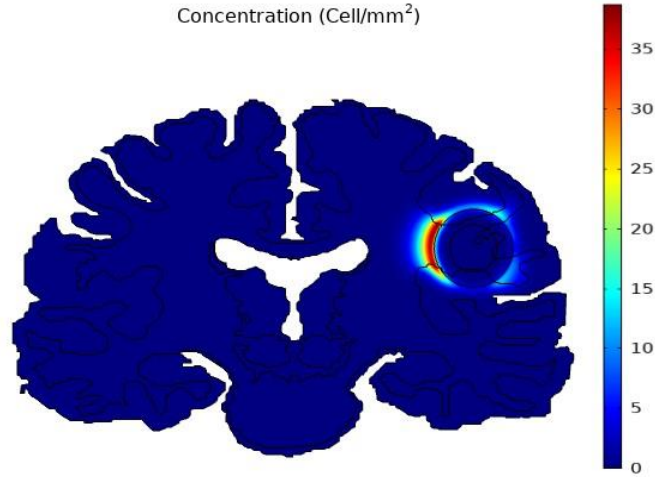

Figure 2. Distribution of cancerous cells after surgery

Treatment Schedule including chemotherapy and radiotherapy after surgery is as follows:

Concurrent chemotherapy using $75 \frac{mg}{m^2}$ temozolomide every day for 6 weeks
Conventional Radiotherapy (Dose = 2 Gy per day) for 6 weeks except for weekends
Adjuvant chemotherapy using $200 \frac{mg}{m^2}$ temozolomide starting 68 days after radiotherapy 5 days a week for 6 courses

It was assumed that due to the high diffusivity of these tumors and high concentrations of cancerous cells on the boundary of the resected area, recurrence of the tumor in the resected area with different diffusion rate, which is the average of the diffusion rate of white and grey matter, was occurred. The concentration of cancerous cells for alpha 0.018, 0.030, and 0.036 were simulated for day 400 (6 months after the last day of therapy) and depicted in Figure 3. Moreover, the concentration of cancerous cells, when no treatments were applied, were simulated for a better understanding of the result.

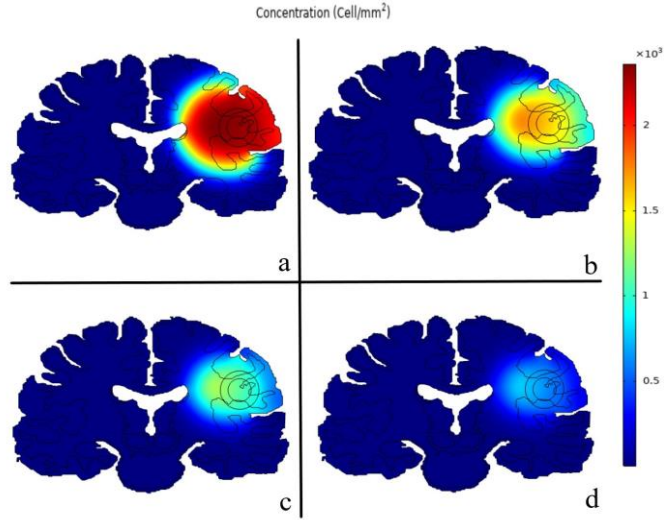

Figure 3. concentration of tumor cells for a) growth without treatment b) $\alpha = 0.018 \frac{1}{Gy}$ c) $\alpha = 0.030 \frac{1}{Gy}$ d) $\alpha = 0.036 \frac{1}{Gy}$

As depicted in figure 3, the concentration of cancerous cells decreased while increasing parameter α, as higher α, is more susceptible to radiotherapy and the tissue absorbs more radio energy. The best result was observed for $\alpha = 0.036 \frac{1}{Gy}$ as there is no high concentration of cancerous cells in the area. The concentration of cancerous cells when $\alpha = 0.030 \frac{1}{Gy}$, is higher in white matter in comparison to grey matter due to the high diffusivity of white matter and the high killing rate of chemotherapy in grey matter. Moreover, the recurrence of tumor in the resected area resulted in a more heterogeneous profile of concentration in the area, which is higher in the vicinity of white matter and lower in the vicinity of grey matter. Furthermore, the location of the tumor is of great importance as the presence of brain physical boundaries and the location of the ventricles contribute to higher cell density in these areas as seen in Figure 3.

The threshold of detection of MRI (T1Gd) was taken to be $400 \frac{cell}{mm^2}$, which means that cancerous cells with concentration below that of the threshold, will not appear in the image. In order to compare the clinical result with the simulated model, it is necessary to know the visible area. The simulated area which appears in the MRI (T1Gd) Image is depicted in Figure 4.

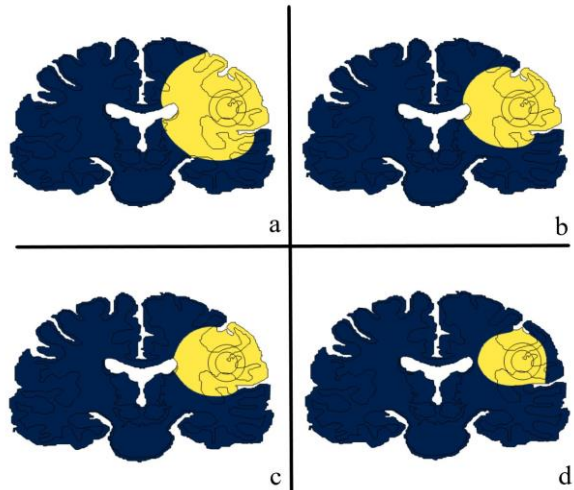

Figure 4. Detectable Area of tumor based on MRI T1Gd threshold of detection for a) growth without treatment b) $\alpha = 0.018 \frac{1}{Gy}$ c) $\alpha = 0.030 \frac{1}{Gy}$ d) $\alpha = 0.036 \frac{1}{Gy}$

Simulating visible area of the tumor is of great importance especially when $\frac{\rho}{D}$ is small because the diffusion coefficient is greater than the proliferation rate and a small part of the tumor is detectable. In all conditions depicted in Figure 4, the patient received the same dosage of radiotherapy, but the extent of difference in the visible area of the tumor for different alphas is different and the best result belongs to alpha 0.036.

The concentration of cancerous cells along a 19 mm line, created toward normal tissue, 6 months after therapy is depicted in Figure 5.

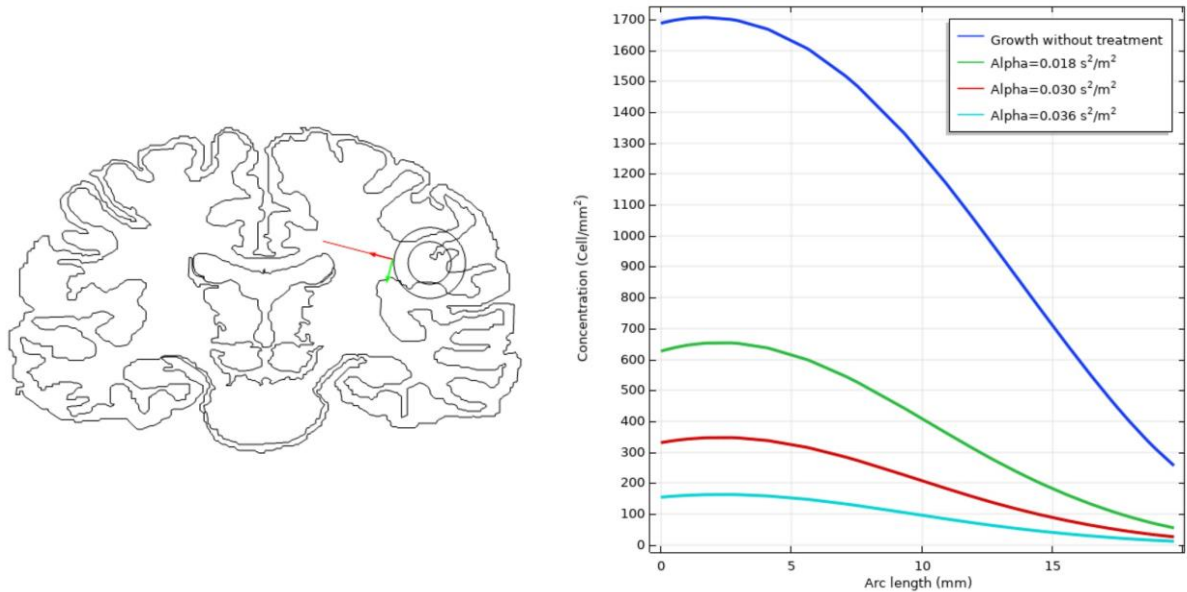

Figure 5. The concentration of cancerous cells along the line toward normal tissue 6 months after therapy for a) growth without treatment b) $\alpha = 0.018 \frac{1}{Gy}$ c) $\alpha = 0.030 \frac{1}{Gy}$ d) $\alpha = 0.036 \frac{1}{Gy}$

The area of the tumor until reaching the radius of death (3 cm) was simulated and depicted in Figure 6.

In the first period of 150 days after surgery, almost all parts of the tumor disappeared. Recurrence appeared after this period which same recurrence happens in real clinical cases. The time taken after surgery for the tumor to reach to its death radius is written in table 2.

Table 2. Survival time of patient for a) growth without treatment b) $\alpha = 0.018 \frac{1}{Gy}$ c) $\alpha = 0.030 \frac{1}{Gy}$ d) $\alpha = 0.036 \frac{1}{Gy}$

| Survival time | $\alpha [\frac{1}{Gy}]$ |
|---|---|
| 331 days | Growth without treatment |
| 388 days | 0.018 |
| 410 days | 0.030 |
| 434 days | 0.036 |

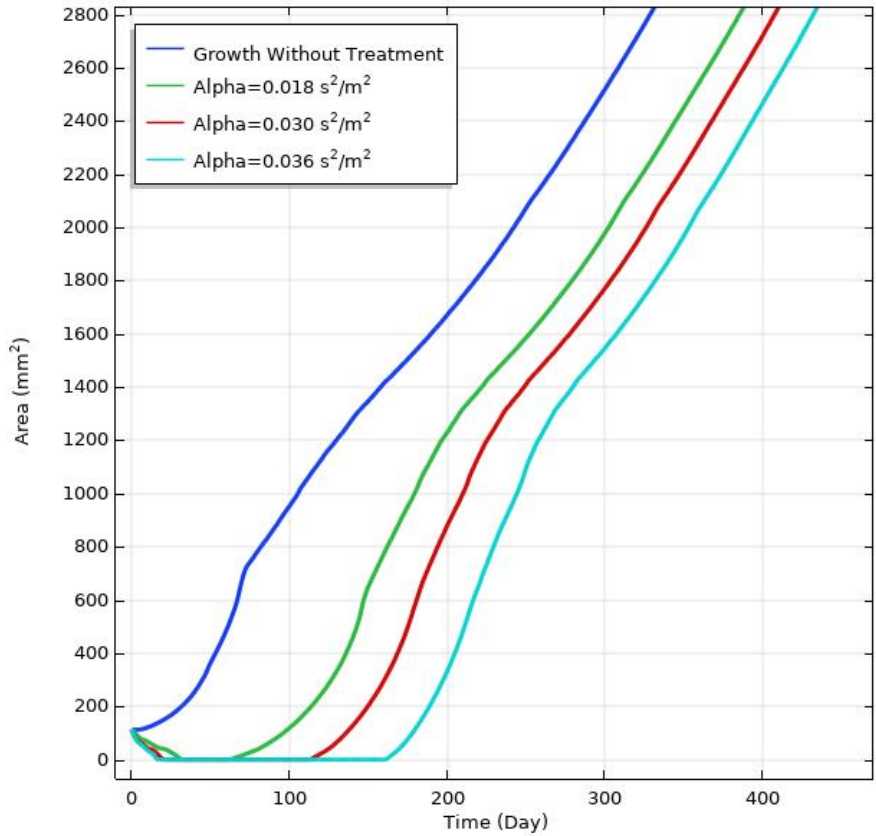
Figure 6. The simulated area of tumor up to the radius of death

As shown in Table 2, the survival time of patient 1 prolonged thanks to the treatment from 331 days to 388,410, and 434 days for $\alpha = 0.018 \frac{1}{Gy}$, $\alpha = 0.030 \frac{1}{Gy}$, and $\alpha = 0.036 \frac{1}{Gy}$ respectively.

The proliferation and diffusion rate of the tumor was slowed down after the patient underwent the treatment. These growth rates almost have been hampered by chemo and radiotherapy, especially in the period of radiotherapy.

## 3.2 Conventional Vs Hyperfractionated Radiotherapy

Hyperfractionated radiotherapy is almost used in order to reduce the toxicity and damage relating to normal tissues around cancerous tissues. To do so, the dosage is given often in two fractions per day instead of giving one fraction per day in conventional radiotherapy. We have investigated two types of common radiotherapy namely conventional and hyperfractionated with dosage and timing for a week written in table 3.

Table 3 Dosage of radiotherapy for 5 days a week

| Type | Day 1 | Day 2 | Day 3 | Day 4 | Day 5 |
|---|---|---|---|---|---|
| Conventional | 2 Gy | 2 Gy | 2 Gy | 2 Gy | 2 Gy |
| Hyperfractionated | 2×1.2 Gy | 2×1.2 Gy | 2×1.2 Gy | 2×1.2 Gy | 2×1.2 Gy |

All in all, six courses of radiotherapy for 5 days a week along with previously mentioned chemotherapy was simulated for the patient. The concentration of cancerous cells 6 months after surgery is depicted in Figure 7.

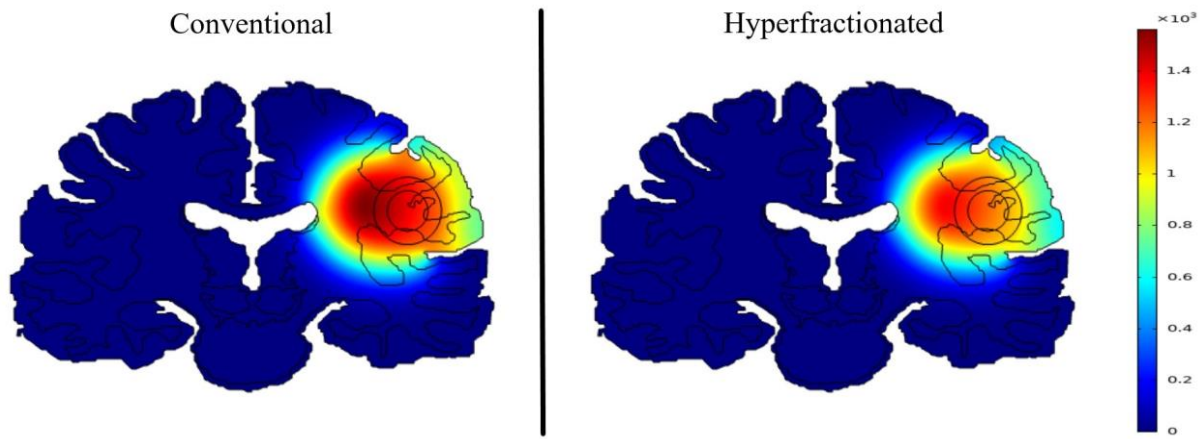

Figure 7. Concentration of cancerous cells 6 months after surgery for conventional and hyperfractionated radiotherapy

As seen in Figure 7, high concentration of cancerous cells occurred in the white matter in the vicinity of resected area as the diffusion rate of white matter is more than the grey matter and chemotherapy effect is lower in white matter. Hyperfractionated radiotherapy as seen in Figure 7 has a better effect in lowering the concentration of cancerous cells as the amount of high concentration of cancerous cells in hyperfractionated radiotherapy is considerably lower than conventional radiotherapy. Moreover, it is worth mentioning that the toxic effect of hyperfractionated radiotherapy on the surrounding normal tissue is lower than conventional radiotherapy which makes the process of treatment more tolerable for the patient with much lesser side effects.

The simulated area of the tumor 6 months after surgery for both types of radiotherapy which appears in the MRI T1Gd Image is depicted in Figure 8.

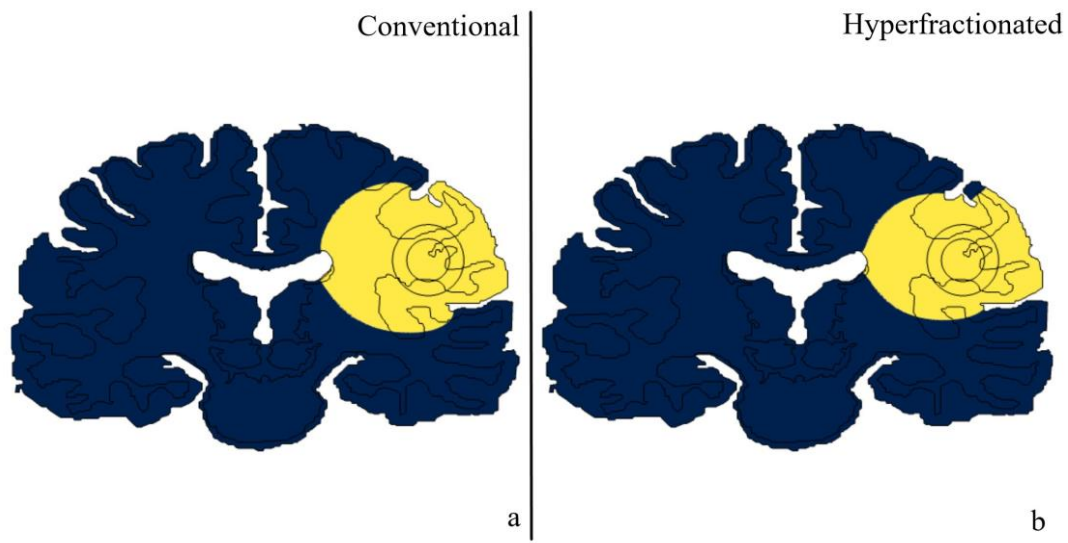

Figure 8. Detectable Area of the tumor 6 months after surgery based on MRI T1Gd threshold of detection for a) conventional radiotherapy b) hyperfractionated radiotherapy

The area of the tumor for both types of radiotherapy until reaching the radius of death (3 cm) were simulated and depicted in figure 8.

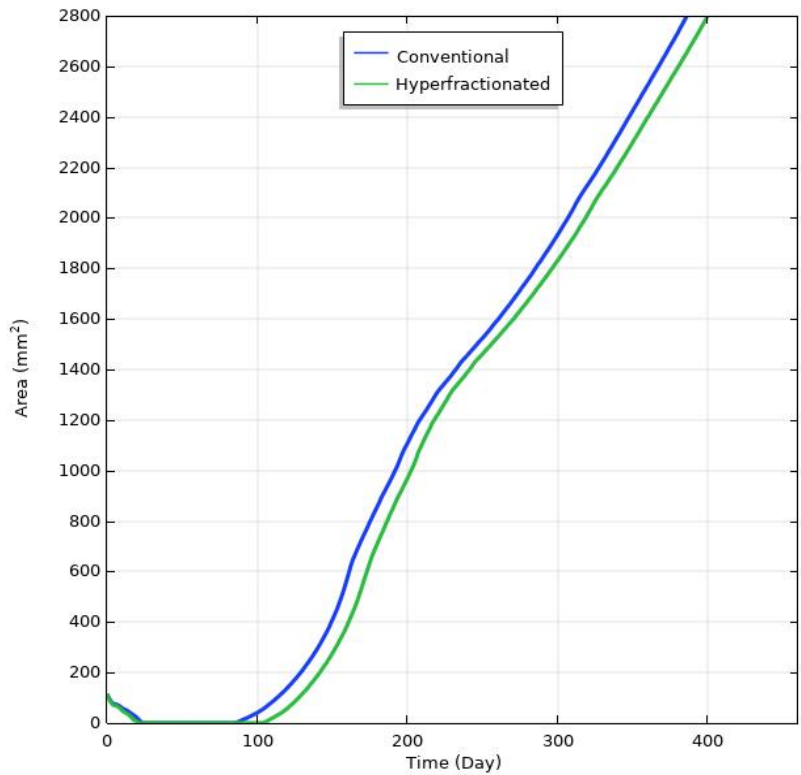

Figure 9. The simulated area of the tumor for both types of radiotherapy up to the radius of death

As seen in Figure 9, hyperfractionated radiotherapy showed a slightly better result than conventional radiotherapy. It is worth mentioning that although they have shown little difference in the area of the tumor, the maximum density of cancerous cells is considerably lower in hyperfractionated radiotherapy as depicted in Figure 9.The survival time reported in the clinical data by Wang was 417 days which shows good agreement with the data simulated.

## 4. Conclusion

In this study, we have modeled heterogeneous brain geometry in order to simulate the growth of tumor and the treatment response of resection plus chemotherapy and radiotherapy more accurately. Temozolomide with different dosages, based on clinical data, has been used in order to preclude the tumor growth and kill the remaining tumor cells along and after radiotherapy. Based on the detection threshold of MRI(T1Gd), the visible area of the tumor was simulated, wishing to provide a clearer vision for medical experts to compare their clinical results with this mathematical modeling. Accurate distribution of cancerous cells for low, medium, and high radiosensitivity parameters have been simulated and survival time and the area of the tumor until death has been calculated. Furthermore, the extent of difference in treatment between conventional and fractionated radiotherapy was shown. The survival time of patient was increased by 2 months and more than 3 months at the lowest and highest sensitivity of radiation therapy, respectively.The result showed hyperfractionated radiotherapy was more effective than conventional radiotherapy in reducing the concentration of cancerous cells,as hyperfractionated method allows administration of higher total dose. The total area covered with cancerous cells had little difference.

## Compliance with Ethical Standards

**Conflict of interests** All of the authors declare that they have no conflict interest.

**Ethical approval** This article does not contain any studies with human participants or animals performed by any of the authors.